\RequirePackage{fix-cm}
\documentclass[smallextended]{svjour3}       
\smartqed  
\usepackage{graphicx}
\journalname{Rendiconti Lincei}
\def \agile{AGILE }

\def \fermi{FERMI }

\def \be {\begin{equation}}
\def \en {\end{equation}}

\begin{document}

\title{A multi-wavelength pipeline for pulsar searches
}


\author{Maura Pilia$^1$
 Alessio Trois$^1$, Matteo Bachetti$^1$, Alberto Pellizzoni$^1$, Giuseppe Atzeni$^2$, Elise Egron$^1$, Maria Noemi Iacolina$^3$, Sara Loru$^{1,2}$, Antonio Poddighe$^1$, Valentina Vacca$^1$}

\authorrunning{M. Pilia et al.} 

\institute{\email: {maura.pilia@inaf.it} \at  $^1$INAF - Osservatorio Astronomico di Cagliari, via della Scienza 5, I-09047 Selargius (Cagliari), Italy \\
  $^2$Dipartimento di Fisica, Universit{\'a} degli Studi di Cagliari, SP Monserrato-Sestu, KM 0.7, 09042 Monserrato, Italy \\
  $^3$ASI - Italian Space Agency - via del Politecnico snc, I-00133 Roma, Italy}

\date{Received: date / Accepted: date}

\maketitle

\begin{abstract}
Pulsar studies in the recent years have shown, more than others, to have benefited from a multi-wavelength approach.
The INAF - Astronomical Observatory in Cagliari (INAF-OAC) is a growing facility with a young group devoted to pulsar and fast transients studies across the electromagnetic spectrum. Taking advantage of this expertise we have worked to provide a suite of multi-wavelength  software and databases for the observations of pulsars and compact Galactic objects at the Sardinia Radio Telescope (SRT, \cite{bolli15}, \cite{pran17}). In turn, radio pulsar observations at SRT will be made available, in a processed format, to gamma-ray searches using \agile\ and \fermi\ gamma-ray satellite and, in a near future, they will be complementary to polarimetric X-ray observations with IXPE.

\keywords{astronomical data bases: miscellaneous, pulsars: general}
\end{abstract}

\section{Introduction}
\label{intro}

The primary aim of this project is to provide a quicklook suite for the observer using SRT for pulsars, compact objects or fast transients studies.
The observer will be provided with a set of online tools to be accessible during the observation at SRT, in order to look in real time for high energy counterparts of the target(s) they are pointing at.
As a demostrator of this package, which will be made available when SRT observations are up to speed later in the year, two offline independent pipelines have already been tested for radio and $\gamma$-rays, and a third one is in the making for the addition of X-ray data.

\begin{figure*}
\begin{center}
  \includegraphics[width=0.9\textwidth]{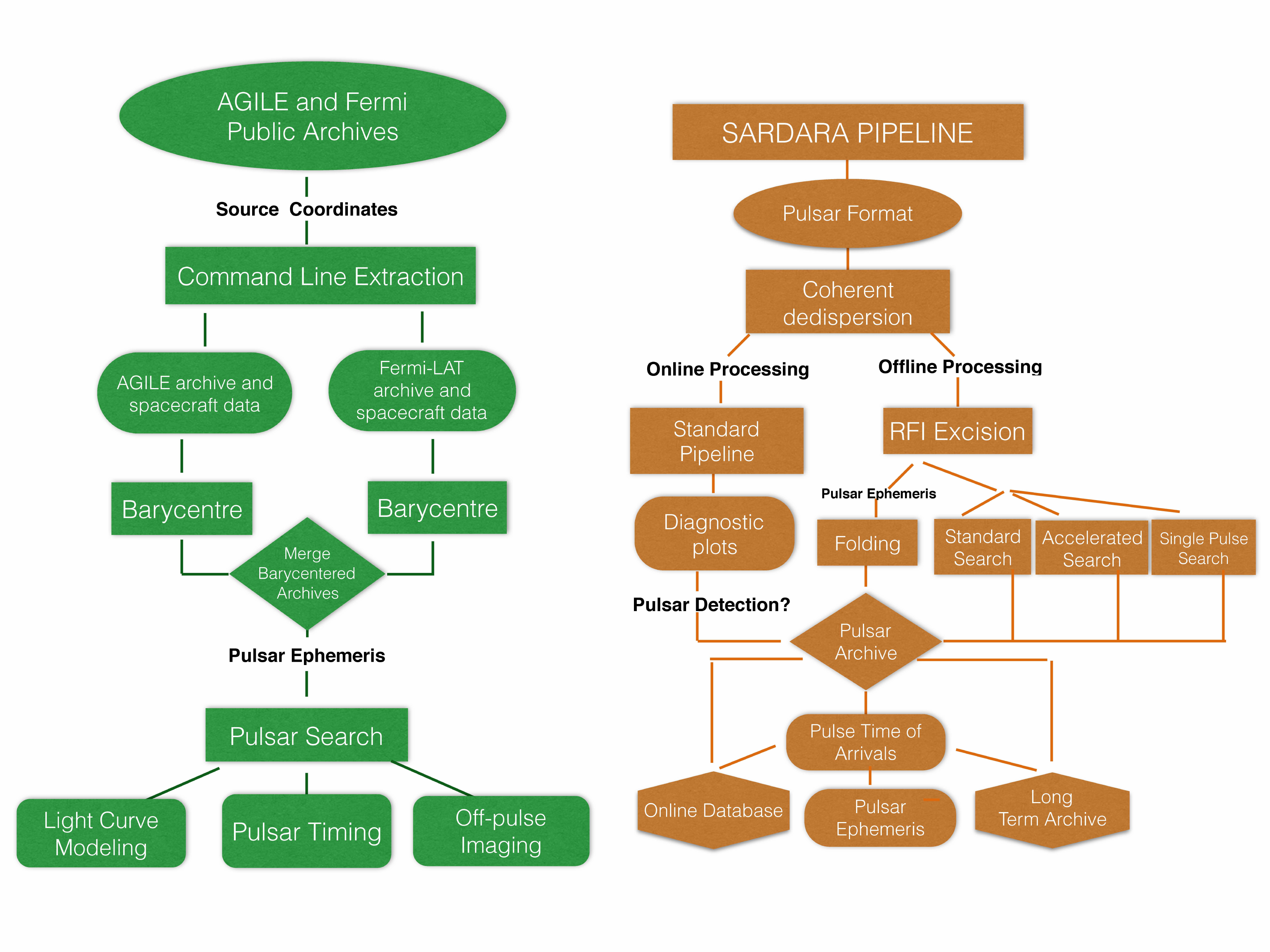}
 \caption{Outline of the radio (left) and $\gamma$-ray (right) pipelines.}
   \label{fig1}
\end{center}
\end{figure*}

\section{Outline of the Infrastructure}
We present a brief description of each pipeline, with the radio and $\gamma$-ray ones schematically illustrated in Figure\,\ref{fig1}. 

{\underline{\it Radio pipeline}}. Observations of pulsars and compact objects at SRT will be processed by the ROACH-2 based SARDARA backend \cite{melis18}. SARDARA will provide full-Stokes observations with different configurations (up to 16000 frequency channels and 16000 spectra per second) to optimally observe pulsars at all bands available at SRT (P-, L-, C- and K-band, soon also a 7-feed S-band) using coherent or partially coherent dedispersion. 
We have developed a python wrapper around the standard pulsar search software {\tt presto} \cite{ransom01} which allows standard and accelerated search plus single pulse search for strong pulsars and fast transients. Alternative search algorithms, based on the use of the Karhunen-Loeve transform instead of the "classical" Fast Fourier Transform, are also under development. The pipeline is being optimised for RFI-excision given the highly polluted interference environment surrounding the telescope, especially at the lowest frequencies. Candidate selection will be partly authomatised through the application of machine learning techniques (either one that seems optimal for SRT data or a combination of more than one in order to avoid the bias of a single classification). Pulsar gating is also implemented, to improve the search for underlying diffuse emission surrounding the compact object (e.g. pulsar wind nebulae, PWNe). 

{\underline{\it Gamma-ray pipeline}}. Online archives are available for $\gamma$-ray observations of the two $\gamma$-ray telescopes in orbit: the AGILE-GRID and Fermi-LAT. We developed a simple command line interface to the archives of the two telescopes, so that data from both can be directly downloaded on our servers for the requested position and time interval. Additional specifics such as energy-range, data quality, off-axis angle, albedo filtering, can be also provided. Given the center position, the pipeline barycenters the data and, if so chosen, it can create a single fits file including the GTIs and barycentered times for the combined dataset. After barycentering is performed, standard folding can be carried out using a known ephemeris of the source (see e.g. \cite{pell09a}), and follow-up off-pulse analysis of nebular emission (as in \cite{pell10}). Optimised tools for exposure calculation for each satellite are being tested, as is blind search on unidentified sources. A Bayesian approach combined with already tested machine learning methods \cite{sazp17} will be used to select candidate Galactic compact objects from $\gamma$-ray unidentified sources.

{\underline{\it X-ray pipeline}}. A similarly structured command-line pipeline has already been developed in-house to download and analyse NuSTAR data and is available upon request (https://gitlab.com/matteobachetti/heasarc\_pipelines).
Because our group is part of the Italian team responsible for the development, calibration and validation of the scientific software for the Imaging X-ray Polarimetry Explorer (IXPE, \cite{wei16}), future prospects include the development of a pipeline for the processing of IXPE public data as an important addition to our high-energy database. IXPE will be able to do polarimetry and timing analyses (time tagging accuracy of less than $100\mu$s) at energies from 2 to 10 keV, thus opening an important new window on pulsars and PWNe.

%
%

\begin{acknowledgements}
The project "Development of a Software Tool for the Study of 
Pulsars from Radio to Gamma-rays using Multi-mission Data" is supported by the Autonomous Region of Sardinia (RAS), CRP-25476.
\end{acknowledgements}



\end{document}